\documentclass[aps,pra,reprint,nofootinbib]{revtex4-2}

\usepackage{amsmath}
\usepackage{amssymb}
\usepackage[hidelinks]{hyperref}

\newcommand{\Pow}{\mathcal{P}}
\newcommand{\Part}{\Pi}
\newcommand{\BA}{\mathcal{B}}
\newcommand{\DSD}{\operatorname{DSD}}
\newcommand{\Span}{\operatorname{span}}
\newcommand{\ran}{\operatorname{ran}}
\newcommand{\dit}{\operatorname{dit}}
\newcommand{\indit}{\operatorname{indit}}

\begin{document}

\title{The Logic of Partitions and Partition Logics: Ore's Correspondence, Contextual Pasting, and Direct-Sum Decompositions}

\author{Karl Svozil}
\email{karl.svozil@tuwien.ac.at}
\affiliation{Institute for Theoretical Physics, TU Wien,
Wiedner Hauptstra\ss{}e 8--10/136, 1040 Vienna, Austria}
\date{\today}

\begin{abstract}
The term ``partition logic'' denotes two constructions at different levels.
In automaton and generalized-urn models, selected partitions generate Boolean
event algebras whose contextwise union forms a concrete pasted event structure;
in Ellerman's framework, whole partitions are classifications governed by
refinement and partition operations.  For a finite set $U$, Ore's
correspondence maps each generator $\pi$ to its Boolean algebra $\BA(\pi)$, but
it neither identifies the pasted carrier with $\Part(U)$ nor makes pasting a
partition operation.  It yields
$\BA(\pi\wedge\sigma)=\BA(\pi)\cap\BA(\sigma)$ and
$\BA(\pi\vee\sigma)=\langle\BA(\pi)\cup\BA(\sigma)\rangle_{\rm BA}$, where
$\langle\cdot\rangle_{\rm BA}$ denotes Boolean-algebra generation.  Thus meet
captures the common event algebra, whereas join gives the ambient Boolean
closure.  Chinese-lantern, Firefly, and triangular examples distinguish shared
events, atomic intertwining, and inherited concrete order.  Ellerman's
direct-sum decompositions (DSDs) provide a vector-space analogue: component
projections of an orthogonal DSD resolve the identity and encode exclusive
outcomes, but its components are not equivalence classes of vectors.  Gleason
and Kochen--Specker applications require globally context-consistent valuations
on those projections.
\end{abstract}

\maketitle

\section{Introduction}

The different uses of the expression ``partition logic'' can be formulated as
a type difference.  A partition may label an experimental question: its
blocks, regarded as subsets of a hidden-state set, support the possible
outcome-event propositions, and unions of its blocks support coarser event
propositions.  Alternatively, the whole partition may itself be the
mathematical object, namely a classification determined by which pairs lie in
the same block and which are distinguished.

Following established usage, we retain ``partition logic'' when referring to
the automaton and generalized-urn literature.  For the construction itself,
``contextual partition pasting'' or ``pasted event structure'' is more literal:
Boolean event algebras are combined along their common events.

The first use occurs in generalized urn models, finite automata, and
contextual event structures
\cite{wright,schaller-95,dvur-pul-svo,svozil-2001-eua}.  A selected family
of partitions gives a selected family of Boolean algebras of events; pasting
these algebras yields a partial Boolean event structure.  Under additional
structural conditions it may form an orthoalgebra, an orthomodular poset, or,
in special cases, an orthomodular lattice.  The second use is Ellerman's logic
of partitions
\cite{ellerman-2010-partition-logic,ellerman-2014-introduction,ellerman-2019-graph-partitions}, which equips the full set of
partitions of a universe with operations taking partitions to partitions.
Ellerman subsequently linearizes the construction by replacing set
partitions with direct-sum decompositions (DSDs) of vector spaces
\cite{ellerman-2018-dsd-quantum-logic}.

The two uses have distinct histories, carriers, and operations.  For finite
$U$, Ore's correspondence identifies partitions of $U$ with Boolean
subalgebras of $\Pow(U)$ \cite{ore-1942-equivalence}.  Rota and collaborators
developed a related logic of commuting equivalence relations
\cite{finberg-mainetti-rota-1996}.  Ore's correspondence supplies an exact
comparison at the level of context generators on a fixed $U$: each selected
  partition $\pi_k$ corresponds to the Boolean context $\BA(\pi_k)$.  In the
  concrete set representation the carrier is then formed as
  $L=\bigcup_k\BA(\pi_k)$.  Since this union is
generally not a Boolean algebra, it is generally not $\BA(\tau)$ for any
partition $\tau$.  The meet and join of the generating partitions nevertheless
summarize, respectively, the common
event algebra and the Boolean closure of the chosen concrete context family.

\begin{table*}[t]
\caption{The two uses of partitions.  Ore's correspondence applies to
the generating partitions and their Boolean context algebras; it does not
identify $\Part(U)$ with the pasted carrier.}
\label{tab:two-levels}
\begin{ruledtabular}
\begin{tabular}{p{0.15\textwidth}p{0.36\textwidth}p{0.36\textwidth}}
Aspect & Whole-partition logic & Contextual partition pastings \\
\hline
Carrier & $\Part(U)$, all partitions of $U$
& Concrete: $\bigcup_k\BA(\pi_k)\subseteq\Pow(U)$; abstract: amalgamated
Boolean blocks \\
Elements & Whole partitions
& Concrete: subset events; abstract: context events modulo prescribed
identifications \\
Order & Refinement, equivalently distinction-set (ditset) inclusion
& Concrete: subset inclusion; abstract: context-generated order \\
Operations & Join, meet, implication, and closure-based operations
& Boolean operations are total locally; ambient operations may leave the
concrete carrier, while abstract cross-context operations are partial \\
Nonclassicality & Nondistributivity of the partition lattice
& Partiality and nondistributivity of the pasted event structure \\
Logical reading & Classification questions or abstract attributes;
same/different distinctions
& Event propositions; subset membership and contextual incidence
\end{tabular}
\end{ruledtabular}
\end{table*}

\section{The two mathematical levels}

\subsection{Contextual partition pastings}

Let $U$ be a finite set and let
\begin{equation}
  \pi=\{B_i\}_{i\in I}
\end{equation}
be a partition of $U$.  In the contextual interpretation, each block $B_i$,
regarded as a subset of $U$, supports the unary event proposition $u\in B_i$
associated with one experimental outcome; it is the set of underlying states
yielding that outcome.  Moreover,
\begin{equation}
  \BA(\pi)=
  \left\{\bigcup_{i\in S}B_i:S\subseteq I\right\}
  \subseteq\Pow(U)
  \label{eq:generated-ba}
\end{equation}
is the Boolean algebra of its decidable events.  For a chosen finite family
$\{\pi_1,\ldots,\pi_m\}$, the carrier of the associated pasted event structure is
\begin{equation}
  L=\bigcup_{k=1}^{m}\BA(\pi_k).
  \label{eq:pasted-carrier}
\end{equation}
  Equation~\eqref{eq:pasted-carrier} defines a concrete set-represented union
  inside one fixed $\Pow(U)$: equal subsets are identified and order is inherited
  from inclusion.  This must be distinguished from an abstract pasting of the
  same Boolean blocks, in which only common elements are identified and order
  and partial operations are generated contextwise.  Ambient inclusion can add
  further cross-context comparabilities, as the triangle-shaped family in
  Sec.~\ref{sec:triangle} shows.  Whether the concrete and abstract
  constructions agree is example-dependent.
  Within each context Boolean operations are total.  Across contexts, ambient
  set operations remain defined in $\Pow(U)$ but may leave $L$, whereas the
  corresponding operations of the abstract pasting are partial.  Pasting is a
  construction of an event structure, not a logical connective acting on two
  partitions.

\subsection{Ellerman's logic of whole partitions}

Ellerman's carrier is $\Part(U)$, the set of all partitions of $U$.  On his
reading, a partition $\pi$ represents a classification question or abstract
attribute.  Its blocks are same-answer or same-value classes, and its
distinction set (ditset) $\dit(\pi)$ consists of the ordered pairs assigned
different answers or values.  Thus its primitive semantic instance is
$(u,u')\in\dit(\pi)$, a pair distinction, rather than $u\in S$, an in/out
assertion about one element.  Partition variables and formulas range over
whole classifications, not over event propositions.  We shall therefore
distinguish a whole-partition question from the event propositions in
$\BA(\pi)$
\cite{ellerman-2010-partition-logic,ellerman-2014-introduction}.

This is the semantic face of the subset--partition duality: subsets, viewed as
subobjects, support unary membership assertions, whereas partitions or
equivalence relations, viewed as quotient objects, support binary assertions
  of sameness or difference.  Ore's
correspondence below is a separate fixed-$U$ order isomorphism; it does not
identify these semantic roles.

Ellerman uses the refinement order
\begin{equation}
  \begin{split}
  \sigma\preceq\pi\quad\Longleftrightarrow\quad
  &\text{each block of $\pi$ is contained}\\
  &\text{in some block of $\sigma$}.
  \end{split}
  \label{eq:refinement}
\end{equation}
Finer partitions lie higher.  The indiscrete partition $0_U=\{U\}$ is the
bottom and the discrete partition
\begin{equation}
  1_U=\bigl\{\{u\}:u\in U\bigr\}
\end{equation}
is the top.  Equivalently, refinement is inclusion of distinction sets:
\begin{equation}
  \sigma\preceq\pi
  \quad\Longleftrightarrow\quad
  \dit(\sigma)\subseteq\dit(\pi).
  \label{eq:dit-refinement}
\end{equation}

The join is the common refinement,
\begin{equation}
  \pi\vee\sigma=
  \{B\cap C:B\in\pi,\ C\in\sigma,\ B\cap C\ne\varnothing\},
  \label{eq:partition-join}
\end{equation}
and the meet is the greatest common coarsening.  Its blocks can be obtained by
taking overlap chains among blocks of $\pi$ and $\sigma$, or equivalently by
taking the reflexive-symmetric-transitive closure of
$\indit(\pi)\cup\indit(\sigma)$, where $\indit(\tau)$ is the indistinction set
of pairs lying in the same block of $\tau$.

This convention reverses the order commonly used for equivalence relations.
With equivalence relations ordered by relation inclusion, the names ``join''
and ``meet'' in Eqs.~\eqref{eq:partition-join} and the preceding description
are exchanged.  Ellerman's convention is deliberate: it places the discrete
partition at the top and makes refinement parallel to inclusion of ditsets.

\subsection{Ore's correspondence at the level of context generators}

For finite $U$, the assignment
\begin{equation}
  \pi\longleftrightarrow\BA(\pi)
  \label{eq:ore-rota-map}
\end{equation}
  is a bijection between partitions of $U$ and Boolean subalgebras of
  $\Pow(U)$.  The finite restriction matters: for infinite $U$, partitions
  correspond not to all Boolean subalgebras of $\Pow(U)$, but to the complete
  atomic Boolean set algebras closed under arbitrary unions of their atoms.  The blocks of
  $\pi$ are the atoms of $\BA(\pi)$.  Semantically,
however, the correspondence does not identify its two sides: $\pi$ is one
whole classification question, whereas $\BA(\pi)$ is a family of
  membership-based event propositions.  Write
 $\langle S\rangle_{\rm BA}$ for the Boolean algebra generated by a family
 $S$.  The following identities are the order-theoretic core of the
 comparison:
\begin{align}
  \sigma\preceq\pi
  &\quad\Longleftrightarrow\quad
  \BA(\sigma)\subseteq\BA(\pi),
  \label{eq:ore-order}\\
  \BA(\pi\vee\sigma)
  &=\left\langle\BA(\pi)\cup\BA(\sigma)\right\rangle_{\rm BA},
  \label{eq:ore-join}\\
  \BA(\pi\wedge\sigma)
  &=\BA(\pi)\cap\BA(\sigma).
  \label{eq:ore-meet}
\end{align}
Indeed, Eq.~\eqref{eq:ore-order} makes the bijection an order isomorphism.
Intersection is the greatest common Boolean subalgebra, and the Boolean
algebra generated by a union is the least common Boolean superalgebra.  This
proves Eqs.~\eqref{eq:ore-join} and \eqref{eq:ore-meet}.  Equivalently, the
atoms of the generated algebra in Eq.~\eqref{eq:ore-join} are the nonempty
intersections of blocks of $\pi$ and $\sigma$, while the atoms of the
intersection algebra in Eq.~\eqref{eq:ore-meet} are the blocks of their
greatest common coarsening.

The two relational descriptions place closure on opposite operations.  On
the Boolean-subalgebra side, meet is literal intersection and join requires
Boolean generation.  On the distinction side,
\begin{align}
  \indit(\pi\wedge\sigma)
  &=\operatorname{cl}_{\rm eq}
    \bigl(\indit(\pi)\cup\indit(\sigma)\bigr),
  \label{eq:meet-equivalence-closure}\\
  \indit(\pi\vee\sigma)
  &=\indit(\pi)\cap\indit(\sigma),\nonumber\\
  \dit(\pi\vee\sigma)
  &=\dit(\pi)\cup\dit(\sigma),
  \label{eq:join-dit-union}
\end{align}
where $\operatorname{cl}_{\rm eq}$ denotes reflexive-symmetric-transitive
closure.  Thus equivalence closure occurs at partition meet, while Boolean
closure occurs at partition join.

There is also a canonical answer-to-question map that should not be confused
with an order or lattice morphism:
\begin{equation}
  \begin{gathered}
  q:\Pow(U)\longrightarrow\Part(U),\\
  q(S)=
    \begin{cases}
      \{S,U\setminus S\},&\varnothing\subsetneq S\subsetneq U,\\
      0_U,&S=\varnothing\text{ or }S=U.
    \end{cases}
  \end{gathered}
  \label{eq:answer-question-map}
\end{equation}
It forgets which complementary answer was called yes:
$q(S)=q(U\setminus S)$.  For every partition $\pi$,
\begin{equation}
  S\in\BA(\pi)
  \quad\Longleftrightarrow\quad
  q(S)\preceq\pi.
  \label{eq:event-binary-question}
\end{equation}
Consequently, if
\begin{equation}
  \mathcal M=\bigcup_k[0_U,\pi_k],\qquad
  \mathcal M_{\leq2}
  =\{\alpha\in\mathcal M:|\alpha|\leq2\},
  \label{eq:selected-partition-cover}
\end{equation}
then the concrete event carrier satisfies
\begin{equation}
  L=q^{-1}(\mathcal M_{\leq2}),
  \qquad
  [0_U,\pi]\cap[0_U,\sigma]=[0_U,\pi\wedge\sigma].
  \label{eq:event-preimage-and-interval-overlap}
\end{equation}
This is an exact bridge between event answers and binary partition questions,
but it does not identify their ordered structures.  If
$\varnothing\subsetneq S\subsetneq T\subsetneq U$, then $S\subset T$ while
the distinct binary partitions $q(S)$ and $q(T)$ are incomparable.  Moreover,
$q$ identifies complementary events.

For a chosen family, Ore's correspondence acts on the context cover, not on
its pasted carrier.  It maps each $\pi_k$ to $\BA(\pi_k)$; in the concrete set
representation one then takes the union of those event algebras, with equal
subsets identified.  Because that union is generally not itself a Boolean
algebra, it is outside the image $\{\BA(\tau):\tau\in\Part(U)\}$; Ore's
correspondence therefore assigns no partition to $L$ and does not turn pasting
into a partition operation.

Equation~\eqref{eq:ore-meet} therefore shows that, for this fixed-$U$
representation, the meet of the generating partitions encodes exactly the
Boolean event algebra shared by two contexts.  In particular,
\begin{equation}
  \pi\wedge\sigma=0_U
  \quad\Longleftrightarrow\quad
  \BA(\pi)\cap\BA(\sigma)=\{\varnothing,U\}.
  \label{eq:no-boolean-overlap}
\end{equation}
Thus $\pi\wedge\sigma\ne0_U$ is equivalent to nontrivial overlap at the
  level of event propositions.  This condition is weaker than atomic
  intertwining in the usual Greechie or Gleason sense.  Atomic intertwining
  requires a block that is an outcome of both contexts,
\begin{equation}
  \pi\cap\sigma\ne\varnothing,
  \label{eq:atomic-intertwining}
  \end{equation}
  where the intersection is taken between $\pi$ and $\sigma$ as sets of blocks;
  in Hilbert space the corresponding condition is a common atomic projection
  (a ray projection in a maximal rank-one context)
  \cite{svozil-2018-b}.

  Here and below, concatenation abbreviates a subset; for example,
  $12=\{1,2\}$.  On $U=\{1,2,3,4,5,6,7,8\}$, for example, let
\begin{equation}
  \pi=\{12,34,56,78\},\qquad
  \sigma=\{13,24,57,68\}.
  \label{eq:overlap-without-shared-atom}
\end{equation}
No block is shared, so $\pi\cap\sigma=\varnothing$, but
\begin{align}
  \pi\wedge\sigma&=\{1234,5678\},\nonumber\\
  \BA(\pi)\cap\BA(\sigma)
  &=\{\varnothing,1234,5678,U\}.
  \label{eq:coarse-common-algebra}
\end{align}
The meet of the generating partitions therefore represents a common
coarse-grained classification even though the contexts have no shared atomic
outcome.

Equation~\eqref{eq:ore-join} identifies the join with the Boolean closure of
the two context algebras inside $\Pow(U)$.  More generally,
\begin{equation}
  \left\langle\bigcup_k\BA(\pi_k)\right\rangle_{\rm BA}
  =\BA\left(\bigvee_k\pi_k\right).
  \label{eq:family-hull}
\end{equation}
The left-hand side is the Boolean algebra generated by $L$, not $L$ itself;
the join describes a closure of the concrete context cover rather than the
pasted event structure.

That closure need not be an invariant of the abstract pasting.  Consider three
binary contexts given by the coordinate partitions of bit strings.  On
$U_4=\{\mathtt{000},\mathtt{011},\mathtt{101},\mathtt{110}\}$, the even-parity
strings, their contextwise union is
the horizontal sum of three four-element Boolean algebras, and their joint
refinement has four atoms, so its Boolean closure has $16$ elements.  On
$U_8=\{0,1\}^3$, the same three coordinate contexts again have an eight-element
horizontal-sum carrier, isomorphic to the one on $U_4$, but their joint
refinement has eight atoms and the Boolean closure has $256$ elements.  Thus
even isomorphic pasted carriers can have nonisomorphic ambient Boolean
closures: the join summarizes the selected concrete realization, not the
abstract pasted carrier alone.

For a concrete set-partition realization, the point valuations form an
order-determining and separating family of two-valued states: each $u\in U$
defines
\begin{equation}
  v_u(S)=
  \begin{cases}
  1,&u\in S,\\
  0,&u\notin S.
  \end{cases}
  \label{eq:point-valuation}
\end{equation}
If $S\nsubseteq T$, some $u\in S\setminus T$ satisfies
$v_u(S)=1>0=v_u(T)$; thus these valuations determine the inherited order and
separate distinct events.  The concrete carrier's inclusion in the Boolean
algebra
of the joint classical refinement is an order embedding.  When the pasted
carrier is a lattice, this inclusion need not preserve its lattice operations;
more generally, it need not preserve cross-context partial operations.  This
concrete subset representation underlies the
mathematical equivalence with automaton and generalized-urn models.

There is a conditional converse.  Let $L$ be a finite orthoalgebra or
orthomodular poset covered by finite Boolean contexts, and let $\mathfrak S$ be
a family of globally defined two-valued states.  Call $\mathfrak S$
\emph{order determining}
when
\begin{equation}
  a\nleq b
  \quad\Longrightarrow\quad
  \text{some }v\in\mathfrak S\text{ has }v(a)=1,
  \ v(b)=0.
  \label{eq:order-determining-states}
\end{equation}
Then
\begin{equation}
  \eta(a)=\{v\in\mathfrak S:v(a)=1\}
  \label{eq:state-set-representation}
\end{equation}
is an order embedding into $\Pow(\mathfrak S)$, and the images of the atoms in
each context partition $\mathfrak S$.  Thus the contexts acquire a faithful
set-partition representation preserving their Boolean operations.  Conversely,
the point valuations of every faithful concrete partition representation are
order determining
\cite{schaller-95,dvur-pul-svo,svozil-2008-ql}.  Pairwise separation of abstract
event labels is weaker than order determination; the triangle example below
gives an explicit witness.  Independently, an order-determining representing
family need not exhaust all abstract two-valued states.

\section{Examples and the level shift}

\subsection{The four-state Chinese lantern}

Let
\begin{equation}
  U=\{1,2,3,4\},\qquad
  \pi=\{12,34\},\qquad
  \sigma=\{13,24\}.
  \label{eq:four-partitions}
\end{equation}
The induced event algebras are
\begin{align}
  \BA(\pi)&=\{\varnothing,12,34,U\},\\
  \BA(\sigma)&=\{\varnothing,13,24,U\}.
\end{align}
Their contextwise union is the six-element event structure traditionally
called the Chinese-lantern logic.  Its four middle elements belong to two
alternative contexts and do not collectively form one partition.
Cross-context Boolean operations must be distinguished from order operations.
Although $12\cap13=\{1\}$ in $\Pow(U)$, the event $\{1\}$ is absent from the
carrier, so the contextual Boolean conjunction of $12$ and $13$ is undefined.
In the underlying six-element inclusion poset, however, their order-theoretic
meet exists and equals $\varnothing$.

At the level of whole partitions,
\begin{align}
  \pi\wedge\sigma&=0_U=\{1234\},
  \label{eq:four-meet}\\
  \pi\vee\sigma&=1_U=\{1,2,3,4\},
  \label{eq:four-join}
\end{align}
where the last expression denotes four singleton blocks.  These equations
neither reproduce nor contradict the pasting; they summarize this concrete
context pair.  By Eqs.~\eqref{eq:no-boolean-overlap} and \eqref{eq:ore-join},
the meet records its trivial overlap algebra, while the join records its full
Boolean closure inside $\Pow(U)$, which distinguishes all four hidden states.

The example also separates two occurrences of non-distributivity.  The set
$\{0_U,\pi,\sigma,1_U\}$ is a four-element Boolean algebra inside
$\Part(U)$, whereas the six-element pasted event structure is
non-distributive.  Conversely, the three $2+2$ partitions
\begin{equation}
  \{12,34\},\qquad\{13,24\},\qquad\{14,23\}
\end{equation}
together with $0_U$ and $1_U$ form the five-element diamond lattice $M_3$ as a
sublattice of $\Part(U)$.  The two
nondistributivities occur at different logical levels.

\subsection{The Firefly logic \texorpdfstring{$L_{12}$}{L12}}

Consider
\begin{equation}
  \pi=\{12,34,5\},\qquad
  \sigma=\{13,24,5\}
  \label{eq:firefly-partitions}
\end{equation}
on $U=\{1,2,3,4,5\}$.  Their event algebras are
\begin{align}
  \BA(\pi)
  &=\{\varnothing,12,34,5,1234,125,345,U\},
  \label{eq:firefly-ba-pi}\\
  \BA(\sigma)
  &=\{\varnothing,13,24,5,1234,135,245,U\}.
  \label{eq:firefly-ba-sigma}
\end{align}
Their common part is
\begin{equation}
  \BA(\pi)\cap\BA(\sigma)
  =\{\varnothing,5,1234,U\}.
  \label{eq:firefly-overlap}
\end{equation}
Consequently their contextwise union has $8+8-4=12$ elements and is the pasted
event structure known as the Firefly logic $L_{12}$.

Ellerman's operations give
\begin{align}
  \pi\wedge\sigma&=\{1234,5\},
  \label{eq:firefly-meet}\\
  \pi\vee\sigma&=1_U.
  \label{eq:firefly-join}
\end{align}
The meet is the partition representing the shared binary question ``$5$ or
not $5$,'' since
\begin{equation}
  \BA(\pi\wedge\sigma)
  =\BA(\pi)\cap\BA(\sigma).
\end{equation}
Here the overlap is also atomic because $\pi\cap\sigma=\{5\}$.  The meet
records the Boolean algebra generated by the shared outcome and its
complement, while the join records the complete hidden-state refinement.  The
twelve event propositions do not become twelve partition elements; their
generating contexts are described one level higher.

\subsection{A three-context stress test: triangular incidence}
\label{sec:triangle}

A cyclic cover with nontrivial pairwise overlaps is obtained on
$U=\{1,2,3,4\}$ from
\begin{equation}
  \begin{aligned}
  \pi_1&=\{1,24,3\},&
  \pi_2&=\{3,14,2\},\\
  \pi_3&=\{2,34,1\}.
  \end{aligned}
  \label{eq:triangle-partitions}
\end{equation}
Their context incidence is triangular: $\pi_1$ and $\pi_2$ share the outcome
block $3$, $\pi_2$ and $\pi_3$ share $2$, and $\pi_3$ and $\pi_1$ share $1$.
Ellerman's pairwise meets are
\begin{equation}
  \begin{aligned}
  \pi_1\wedge\pi_2&=\{3,124\},\\
  \pi_2\wedge\pi_3&=\{2,134\},\\
  \pi_3\wedge\pi_1&=\{1,234\},
  \end{aligned}
  \label{eq:triangle-meets}
\end{equation}
whereas
\begin{equation}
  \pi_i\vee\pi_j=1_U\quad(i\ne j),
  \qquad
  \bigwedge_{i=1}^{3}\pi_i=0_U.
  \label{eq:triangle-extremes}
\end{equation}
By Eq.~\eqref{eq:ore-meet}, the pairwise context intersections are the
four-element Boolean algebras generated by the three binary partitions in
Eq.~\eqref{eq:triangle-meets}, while the triple intersection is
$\{\varnothing,U\}$.  Writing $\BA_i=\BA(\pi_i)$, inclusion--exclusion gives
the concrete set-represented carrier
\begin{equation}
  \begin{aligned}
  L_{\triangle}^{\rm conc}
  &=\BA_1\cup\BA_2\cup\BA_3\\
  &=\Pow(U)\setminus\bigl\{\{4\},\{1,2,3\}\bigr\},\\
  |L_{\triangle}^{\rm conc}|&=14.
  \end{aligned}
  \label{eq:triangle-carrier}
\end{equation}

This concrete ordered carrier is not the abstract Boolean-block amalgam
determined only by those context overlaps.  Inherited inclusion adds, for
example,
\begin{equation}
  \{1\}\subset\{1,4\},\qquad
  \{2\}\subset\{2,4\},\qquad
  \{3\}\subset\{3,4\},
  \label{eq:triangle-extra-order}
\end{equation}
although, in each displayed inclusion, the two events belong to different
contexts.  Thus $14$, $24$,
and $34$ are context atoms but not global atoms of the concrete inclusion
order.  With $x^\perp=U\setminus x$, that order is not orthomodular: for
$x=1\subset y=14$, the event $4$ is absent, so
$y\wedge x^\perp=14\wedge234=\varnothing$ and
$x\vee(y\wedge x^\perp)=x\ne y$.

The three one-outcome-per-context equations have exactly four solutions,
namely the point valuations $v_u$ of Eq.~\eqref{eq:point-valuation}; they
therefore exhaust the two-valued states of the abstract amalgam.  Every local
outcome is selected by at least one of them, and together they distinguish all
$14$ event labels.  Nevertheless, they are not order determining for the
abstract order: the events $1$ and $14$ are incomparable there, yet
\begin{equation}
  \eta(1)=\{v_1\}\subset\{v_1,v_4\}=\eta(14),
  \label{eq:triangle-state-order-failure}
\end{equation}
Thus the state-set representation produces precisely the extra inclusion in
Eq.~\eqref{eq:triangle-extra-order}.  For the resulting inherited subset order,
the same point valuations are order determining.  Cyclic atomic intertwining
and nontrivial pairwise meets do not by themselves produce state scarcity.

\section{Additional operations in Ellerman's framework}

\subsection{Implication}

Ellerman's implication has no counterpart as an operation of contextual
pasting: that construction defines no operation taking two context partitions
to a partition.  Ellerman defines
$\sigma\Rightarrow\pi$ by examining every block $B$ of $\pi$: if $B$ is
contained in a block of $\sigma$, $B$ is discretized; otherwise it remains
whole.  The operation satisfies
\begin{equation}
  \sigma\Rightarrow\pi=1_U
  \quad\Longleftrightarrow\quad
  \sigma\preceq\pi.
  \label{eq:partition-implication}
\end{equation}
If the partitions arise from random variables or experiments, the discrete
parts of $\sigma\Rightarrow\pi$ mark the $\pi$-outcomes that already determine
the $\sigma$-outcome.  This expresses deterministic functional dependence of
the $\sigma$-outcome on the $\pi$-outcome and is close in spirit to automaton-style
state identification.  In a fixed hidden-state
representation, it may provide a way to organize implication chains across
the intertwined contexts of Hardy-type arguments; establishing such an
application requires additional structure beyond Ore's correspondence
\cite{svozil-2020-hardy}.

\subsection{Boolean-indexed operations and their limitation}

Ellerman's graph method associates a partition operation with any Boolean
truth function.  Edges of the complete graph on $U$ are labeled true or false
according to whether their endpoints are distinguished by each input
partition; edges on which the truth function evaluates false are retained,
and connected components determine the output partition
\cite{ellerman-2019-graph-partitions}.  Equivalently, a relation built from dits and indits is
closed reflexively, symmetrically, and transitively.

The construction differs from a compositional copy of Boolean logic.  The
closure step means that composing the resulting
partition operations need not reproduce composition of their indexing truth
functions.  Ellerman notes, for example, that a compound operation need not
reduce to one of the sixteen basic binary operations, and Boolean tautologies
such as Peirce's law can fail as partition tautologies.

Ellerman's implication also supplies negations.  For every fixed
$\pi\in\Part(U)$, the relative $\pi$-negation
\begin{equation}
  \neg_{\pi}\sigma:=\sigma\Rightarrow\pi,
  \label{eq:relative-negation}
\end{equation}
is defined for every $\sigma\in\Part(U)$.  The $\pi$-regular elements
$x=\sigma\Rightarrow\pi$ form a Boolean core in $[\pi,1_U]$, with bottom
$\pi$ and top $1_U$.
Within that core the complement of $x$ is
\begin{equation}
  \neg_{\pi}x=x\Rightarrow\pi
  =(\sigma\Rightarrow\pi)\Rightarrow\pi.
  \label{eq:boolean-core-complement}
\end{equation}
The special choice $\pi=0_U$ gives Ellerman's globally defined partition
negation
\begin{equation}
  \neg\sigma:=\sigma\Rightarrow0_U=
  \begin{cases}
    1_U,&\sigma=0_U,\\
    0_U,&\sigma\ne0_U.
  \end{cases}
  \label{eq:global-partition-negation}
\end{equation}
Its Boolean core is the two-element algebra $\{0_U,1_U\}$.  It is therefore a
Boolean complement on that core, but not a lattice complement for an
intermediate partition $\sigma$, since
$\sigma\vee\neg\sigma=\sigma\ne1_U$.  This distinguishes a negation defined
on all of $\Part(U)$ from Boolean complementation on the whole partition
lattice
\cite{ellerman-2010-partition-logic,ellerman-2014-introduction}.  The graph
method is a truth-table-indexed definition scheme whose closure operation
creates non-Boolean behavior.

\section{Direct-sum decompositions and contextuality}

\subsection{The partial partition algebra of DSDs}

Throughout this section $V$ is finite dimensional.  Ellerman's vector-space
analogue of a partition is a DSD
\begin{equation}
  \pi=\{V_i\}_{i\in I},\qquad V=\bigoplus_{i\in I}V_i.
\end{equation}
When it arises from a diagonalizable operator, the DSD records the eigenspace
classification while abstracting from the numerical eigenvalues.
The set $\DSD(V)$ is a meet-semilattice under refinement
\cite{ellerman-2018-dsd-quantum-logic}.  Its bottom element is the indiscrete
DSD $0_V=\{V\}$.  A maximal DSD
\begin{equation}
  \omega=\{L_1,\ldots,L_n\}
\end{equation}
consists of one-dimensional subspaces and determines a local partition
algebra
\begin{equation}
  \mathcal I(\omega):=[0_V,\omega]\subseteq\DSD(V),
  \qquad
  \mathcal I(\omega)\cong\Part(\{1,\ldots,n\}).
  \label{eq:local-partition-algebra}
\end{equation}
The isomorphism groups the component lines according to a partition of their
index set.  Every DSD can be refined by choosing bases in its components, so
\begin{equation}
  \DSD(V)=
  \bigcup_{\substack{\omega\in\DSD(V)\\ \omega\ \mathrm{maximal}}}
  \mathcal I(\omega).
  \label{eq:dsd-interval-cover}
\end{equation}
Two DSDs $\{V_i\}$ and $\{W_j\}$ of the same ambient vector space $V$ are
compatible if their nonzero intersections
\begin{equation}
  \{V_i\cap W_j:V_i\cap W_j\ne\{0\}\}
  \label{eq:dsd-protojoin}
\end{equation}
span $V$.  Only then is their join defined.  The absence of a common upper
bound is the DSD manifestation of incompatibility.

For a selected DSD family $\{\alpha_k\}$, the interval union
\begin{equation}
  \mathcal M_{\rm DSD}:=\bigcup_k[0_V,\alpha_k]
  \label{eq:selected-dsd-cover}
\end{equation}
has the same form as the selected set-side cover $\mathcal M$ in
Eq.~\eqref{eq:selected-partition-cover}.  Under literal linearization,
$[0_U,\pi_k]\cong[0_V,\widehat\pi_k]$.  This interval analogy concerns the
local-to-global architecture; it does not identify the local event carriers.
The intervals are generally non-Boolean and overlap through common
coarsenings, whereas projection-event contexts overlap through common
propositions.  The full cover in Eq.~\eqref{eq:dsd-interval-cover} is broader:
literal linearizations of partitions of one $U$ all lie in the single
coordinate interval $\mathcal I(\omega_U)$, while $\DSD(V)$ contains
incompatible maximal DSDs.

Harding's related decomposition construction retains ordered complementary
factor pairs, with orthocomplementation given by interchanging the two factors
\cite{harding-96}.  An unordered binary DSD instead packages complementary
answers into one decomposition.  In the Hilbert-space specialization,
orthogonality selects $M$ and $M^\perp$ and yields self-adjoint projection
events.

\subsection{Orthogonal vector-space partitions and exclusivity}

A finite-dimensional projective measurement, equivalently the spectral family
of a self-adjoint observable, has a precise orthogonal-DSD description.  It is
a maximal context when all its nonzero spectral projections have rank one; a
degenerate spectral family is a coarser projective measurement.  For a
self-adjoint observable with spectral resolution
\begin{equation}
  A=\sum_i a_iP_i,
  \label{eq:spectral-resolution}
\end{equation}
put $V_i=\ran(P_i)$, the range of $P_i$.  Its spectral family satisfies
\begin{equation}
  \mathcal H=\bigoplus_i V_i,
  \qquad
  P_iP_j=0\ (i\ne j),
  \qquad
  \sum_iP_i=I.
  \label{eq:orthogonal-partition-identity}
\end{equation}
Thus $\{V_i\}$ is an orthogonal DSD and $\{P_i\}$ is a partition of the
identity.  Pairwise orthogonality encodes exclusivity of distinct outcomes,
while the resolution of the identity encodes completeness.  More generally,
an orthogonal DSD $\alpha=\{V_i\}_{i\in I}$ determines the Boolean algebra
\begin{equation}
  \mathcal C_\alpha=
  \left\{\sum_{i\in S}P_i:S\subseteq I\right\}
  \label{eq:projection-context}
\end{equation}
of projection events.  For finite orthogonal DSDs $\alpha$ and $\beta$,
\begin{equation}
  \alpha\preceq\beta
  \Longleftrightarrow
  \mathcal C_\alpha\subseteq\mathcal C_\beta,
  \qquad
  \mathcal C_{\alpha\wedge\beta}
  =\mathcal C_\alpha\cap\mathcal C_\beta.
  \label{eq:orthogonal-ore-correspondence}
\end{equation}
Indeed, a nontrivial projection $P$ belongs to $\mathcal C_\alpha$ precisely
when the binary DSD $\{\ran(P),\ran(I-P)\}$ is a coarsening of $\alpha$; the
endpoint projections correspond to $0_V$.  The second identity then follows
from the universal property of the meet.  Projection Boolean algebras may be
pasted along all common projections, including sums of component projections.
This is the usual projection-event construction and
is distinct from the partial partition algebra $\DSD(V)$
\cite{ellerman-2018-dsd-quantum-logic,svozil-2018-b}.

It is worth stressing that an orthogonal DSD is not obtained by taking
equivalence classes under vector orthogonality.  Orthogonality is not reflexive
on nonzero vectors, and it is not transitive.  In $\mathbb C^3$, for example,
\begin{equation}
  e_1\perp e_2,\qquad
  e_2\perp(e_1+e_3),\qquad
  e_1\not\perp(e_1+e_3).
  \label{eq:orthogonality-not-equivalence}
\end{equation}
Nor do the subspaces in Eq.~\eqref{eq:orthogonal-partition-identity} partition
the underlying set of vectors.  A generic vector has a unique component
decomposition $v=\sum_i v_i$ with $v_i\in V_i$, but it need not belong to any
single $V_i$.  Unique decomposition replaces the unique block membership of
a set partition.

After choosing an orthonormal basis adapted to one DSD, its basis rays can be
grouped into a genuine set partition whose block spans are the $V_i$.
Incompatible DSDs, however, need not possess a common adapted basis.  If an
entire family is refined by one coordinate DSD, all of its members are
simultaneously diagonalizable and the construction reduces to the classical
case described below.

Orthogonality can nevertheless serve as the primitive relation through
biorthogonal closure rather than equivalence closure.  For $S\subseteq
\mathcal H$, define
\begin{equation}
  S^\perp=\{x\in\mathcal H:\langle x,s\rangle=0
  \text{ for every }s\in S\}.
  \label{eq:orthogonal-complement-set}
\end{equation}
The fixed points of $S\mapsto S^{\perp\perp}$ are the closed linear
subspaces that form the usual projection logic.  Equivalently, one may use an
orthogonality hypergraph whose vertices are rays (or, more generally,
projections) and whose contexts are maximal orthogonal families resolving the
identity.  The resulting statement is therefore that quantum logic is
generated by orthogonal decompositions, not that orthogonality partitions all
vectors into equivalence classes.

\subsection{Why set-partition models remain classical}

The literal linearization of a finite set $U$ takes a vector space with basis
$\{e_u:u\in U\}$ and maps
\begin{equation}
  \pi=\{B_i\}
  \quad\longmapsto\quad
  \widehat\pi=
  \left\{\Span\{e_u:u\in B_i\}\right\}_i.
  \label{eq:literal-linearization}
\end{equation}
Every literal linearization of a partition of this fixed $U$ is refined by the
same maximal coordinate DSD
\begin{equation}
  \omega_U=\{\Span(e_u):u\in U\}.
\end{equation}
Hence all literal linearizations of partitions on the same finite set $U$ are
mutually compatible; their DSD joins are the literal linearizations of the
corresponding joint set refinements $\bigvee_k\pi_k$.  Equivalently, all lie
below one maximal coordinate DSD.  This reflects the fact that automaton and
urn partition logics are concrete classical models with separating two-valued
states.

For example, literal linearization of Eq.~\eqref{eq:firefly-partitions} gives
\begin{align}
  \widehat\pi
  &=\{\Span(e_1,e_2),\Span(e_3,e_4),\Span(e_5)\},\\
  \widehat\sigma
  &=\{\Span(e_1,e_3),\Span(e_2,e_4),\Span(e_5)\}.
\end{align}
Their join is $\omega_U$, so they can be represented by commuting diagonal
operators.  The operational Firefly event structure does not include that
singleton refinement as an available joint experiment.

The triangle family in Eq.~\eqref{eq:triangle-partitions} makes this
compatibility issue especially explicit.  Put $L_j=\Span(e_j)$ and
$V_{ij}=L_i\oplus L_j$.  Its literal linearizations in $\mathbb C^4$ are
\begin{equation}
  \begin{aligned}
  \widehat\pi_1&=\{L_1,V_{24},L_3\},\\
  \widehat\pi_2&=\{L_3,V_{14},L_2\},\\
  \widehat\pi_3&=\{L_2,V_{34},L_1\}.
  \end{aligned}
  \label{eq:triangle-linearizations}
\end{equation}
These are complete orthogonal three-outcome DSDs of rank pattern $(1,2,1)$,
not maximal rank-one contexts.  All are coarsenings of the same coordinate DSD
$\omega_U=\{L_1,L_2,L_3,L_4\}$, every pair has join $\omega_U$, and their
component projections are diagonal and mutually commuting.  Thus cyclically shared outcomes do not
by themselves yield quantum incompatibility.

Nor does the corresponding six-vertex incidence hypergraph have a faithful
realization by three distinct maximal orthogonal triples in $\mathbb C^3$.
The three shared rays occur pairwise in contexts and therefore form an
orthogonal basis; the private ray in each triple is then forced to coincide
with the remaining shared ray.  Ellerman's unrestricted nonorthogonal DSDs can
realize the incidence by distinct bases, but those bases are not quantum
contexts.  A maximal orthogonal realization in higher dimension requires
refining the private coarse outcomes and therefore represents a different
hypergraph.

\subsection{Intertwining contexts in the Gleason sense}

In the Gleason and Kochen--Specker setting, a context is a maximal Boolean
algebra of commuting projections; equivalently, its atoms form a maximal
pairwise-orthogonal family resolving the identity
\cite{Gleason,kochen1}.  If two orthonormal contexts share a ray $L$,
\begin{align}
  \omega&=\{L,L_2,\ldots,L_n\},\\
  \omega'&=\{L,L'_2,\ldots,L'_n\},
\end{align}
then the binary DSD
\begin{equation}
  \alpha_L=\{L,L^\perp\}
  \label{eq:shared-dsd}
\end{equation}
is refined by both contexts and therefore belongs to
$\mathcal I(\omega)\cap\mathcal I(\omega')$.  In this sense, Ellerman's DSD structure
can encode an observable occurring in several contexts.

The common event need not be atomic.  In $\mathbb C^4$, write
$L_i=\Span(e_i)$ and
$K_{ij}^{\pm}=\Span((e_i\pm e_j)/\sqrt2)$.  Put
\begin{align}
  \gamma&=\{L_1,L_2,L_3,L_4\},\nonumber\\
  \gamma'&=\{K_{12}^{+},K_{12}^{-},K_{34}^{+},K_{34}^{-}\}.
  \label{eq:shared-projection-no-ray}
\end{align}
The contexts share no ray.  With $M_{ij}=\Span(e_i,e_j)$ and $P_M$ denoting
the orthogonal projection onto $M$,
\begin{equation}
  \gamma\wedge\gamma'=\{M_{12},M_{34}\},
  \qquad
  \mathcal C_\gamma\cap\mathcal C_{\gamma'}
  =\{0,P_{M_{12}},P_{M_{34}},I\}.
  \label{eq:coarse-projection-overlap}
\end{equation}
Thus DSD meets detect shared coarse projections as well as shared rays;
atomic intertwining is the stronger condition.

A Hilbert-space incidence model of $L_{12}$ can therefore be made in
$\mathcal H=\mathbb C^3$.  Let $L_5=\Span(e_3)$ and choose
\begin{align}
  \omega&=\{\Span(e_1),\Span(e_2),L_5\},\\
  \omega'&=\left\{
  \Span\!\left(\frac{e_1+e_2}{\sqrt2}\right),
  \Span\!\left(\frac{e_1-e_2}{\sqrt2}\right),L_5\right\}.
  \label{eq:firefly-hilbert-contexts}
\end{align}
These maximal orthogonal DSDs satisfy
\begin{equation}
  \omega\wedge\omega'=\{L_5,L_5^\perp\},
  \qquad
  \omega\vee\omega'\ \text{is undefined}.
  \label{eq:hilbert-firefly-meet}
\end{equation}
Here the protojoin contains only the shared ray $L_5$ and therefore does not
span $\mathcal H$.  More generally, two maximal one-dimensional DSDs have a join only
when they are equal: their nonzero line intersections span $\mathcal H$ only when they
share a complete basis.  Hence in the Firefly example the nontrivial meet
carries the shared incidence information, while the undefined join records
that the maximal contexts are distinct and incompatible.
They thus reproduce the incidence pattern of two incompatible contexts with a
shared outcome.  This is not the literal linearization of the five hidden
states; it retains only the contextual incidence pattern.  Constructing such
graph-incidence realizations by faithful orthogonal representations is studied
systematically in Ref.~\cite{svozil-2018-b}.

\subsection{Additional structure for quantum-foundational applications}

Ellerman defines DSDs for arbitrary direct sums, not necessarily the
orthogonal partitions of the identity in
Eq.~\eqref{eq:orthogonal-partition-identity}.  His partial order and partition
operations do not by themselves encode the orthogonality relation on rays
that carries Gleason and Kochen--Specker arguments.  To address those problems
one must at least restrict to orthogonal DSDs and retain their component
projections.

There is also a level mismatch.  The DSD $\{L,L^\perp\}$ packages the two
complementary events $P_L$ and $I-P_L$ into one whole measurement.  In a
projection logic they are separate propositions.  A Gleason measure requires
a context-independent assignment
\begin{equation}
  \mu(P)\geq0,
  \qquad
  \sum_{L_i\in\omega}\mu(P_{L_i})=1,
  \label{eq:gleason-measure}
\end{equation}
with the same $\mu(P)$ whenever $P$ occurs in several contexts.  For a
finite-dimensional real or complex Hilbert space of dimension at least three,
every normalized measure on all projections that is finitely additive on
mutually orthogonal families has the form
\begin{equation}
  \mu(P)=\operatorname{Tr}(\rho P)
\end{equation}
for a unique $\rho\geq0$ with $\operatorname{Tr}\rho=1$, where
$\operatorname{Tr}$ denotes the trace.  In finite dimension, countable additivity adds no further
requirement.
This probability assignment lives on the
component projections, not on whole DSDs alone.

Indeed, each orthogonal maximal DSD $\omega$ determines the ordinary Boolean
projection algebra $\mathcal C_\omega$ of Eq.~\eqref{eq:projection-context}.
For the two contexts in Eq.~\eqref{eq:firefly-hilbert-contexts},
\begin{equation}
  \mathcal C_\omega\cap\mathcal C_{\omega'}
  =\{0,P_5,I-P_5,I\},
\end{equation}
so their pasted union has $8+8-4=12$ elements and is isomorphic to
$L_{12}$.  This recovery uses the usual event structure obtained by pasting
projection contexts.  The DSD theory organizes the measurements one level
higher; it does not replace the orthogonality and probability structures
needed for quantum foundations.

\subsection{Scarcity of two-valued states}

A separate issue concerns states rather than partition operations.
Ellerman's native carrier consists of whole DSDs, whereas a
Kochen--Specker two-valued state assigns values to their component
projections.  A homomorphism taking whole partitions to $0$ or $1$ would be a
different semantics from a dispersion-free assignment of exactly one outcome
in each measurement context.

The required state layer can be adjoined as follows.  Let
$\Omega$ be a finite family of maximal orthogonal DSDs and let
$A(\Omega)=\bigcup_{\omega\in\Omega}\omega$ be the set of their component
rays, with a shared ray represented by the same element in every context.  An
atomic two-valued assignment is a map satisfying
\begin{equation}
  v:A(\Omega)\longrightarrow\{0,1\},
  \qquad
  \sum_{L\in\omega}v(L)=1
  \quad(\omega\in\Omega).
  \label{eq:two-valued-dsd-state}
\end{equation}
Within each context it extends uniquely by finite additivity:
\begin{equation}
  \widetilde v_\omega\!\left(\sum_{L\in S}P_L\right)
  =\sum_{L\in S}v(L),
  \qquad S\subseteq\omega.
  \label{eq:contextwise-state-extension}
\end{equation}
It defines a state on the pasted projection-event carrier only when these
contextwise extensions agree on every projection in
$\mathcal C_\omega\cap\mathcal C_{\omega'}$.  Equation~\eqref{eq:two-valued-dsd-state}
alone ensures this agreement when every overlap is generated by shared rays,
as in the Firefly example; otherwise agreement on shared non-atomic projections
is an additional global constraint.  For Eqs.~\eqref{eq:shared-projection-no-ray}
and \eqref{eq:coarse-projection-overlap}, for example, selecting $\Span(e_1)$
in $\gamma$ and a ray in $M_{34}$ in $\gamma'$ satisfies both local
one-ray equations but assigns respectively $1$ and $0$ to $P_{M_{12}}$.
Write $\mathcal V(\Omega)$ for the set of
context-consistent assignments.  At the atomic level,
this family is \emph{unital} if every $L\in A(\Omega)$ receives value $1$ in
some $v\in\mathcal V(\Omega)$, and \emph{separating} if every pair of distinct
rays is distinguished by some $v$.  Failure of the first property is
nonunitality; failure of the second is nonseparability.  At the limiting
Kochen--Specker case,
\begin{equation}
  \mathcal V(\Omega)=\varnothing,
  \label{eq:ks-empty-state-space}
\end{equation}
because no context-independent $0$--$1$ assignment obeys all of the
one-outcome-per-context constraints \cite{kochen1,svozil-tkadlec}.

The Firefly logic provides a control example.  For
$\omega=\{L_1,L_2,L_5\}$ and $\omega'=\{L_3,L_4,L_5\}$, put
$\Omega_{12}=\{\omega,\omega'\}$.  Equation~\eqref{eq:two-valued-dsd-state}
has
\begin{equation}
  |\mathcal V(\Omega_{12})|=1+2\mathbin{\cdot}2=5.
  \label{eq:firefly-state-count}
\end{equation}
One state selects the shared ray $L_5$; the other four set $v(L_5)=0$ and
select one ray from each of the remaining pairs.  Hence $L_{12}$ is both
unital and separating despite its intertwined contexts.  Intertwining and
incompatibility therefore do not by themselves imply state scarcity.

For a literal set-partition model, the point valuations in
Eq.~\eqref{eq:point-valuation} are automatically unital and separate all
events.  Consequently, a nonunital, nonseparating, or Kochen--Specker event
structure cannot have a faithful concrete representation of that kind.  Nor
can scarcity be inferred merely from pairwise DSD meets or undefined joins:
it is a global consistency property of the entire orthogonality hypergraph.
The general $\DSD(V)$ framework represents this hierarchy only after
restriction to orthogonal DSDs and imposition of the global constraint
Eq.~\eqref{eq:two-valued-dsd-state} together with the overlap-consistency
condition following Eq.~\eqref{eq:contextwise-state-extension}.  The DSDs
in this orthogonal specialization still organize complete measurement
contexts one level higher, while state scarcity is a global property of
assignments on the pasted projection event structure.

\section{Structural comparison and prospects}

The fixed-$U$ comparison has three exact components.  Ore's correspondence
maps each whole-partition generator $\pi_k$ to its Boolean event context
$\BA(\pi_k)$; the map $q$ in Eq.~\eqref{eq:answer-question-map} sends an
event proposition to its binary partition question; and the interval cover
$\mathcal M$ retains all coarsenings of the selected questions.  None of these
constructions turns the union of event contexts into an operation on
$\Part(U)$.  The linear analogue has the same interval architecture, while
Eq.~\eqref{eq:orthogonal-ore-correspondence} states that the projection algebra
of the greatest common orthogonal-DSD coarsening is exactly the intersection
of the two projection-event algebras.

Other whole-partition quantities require a separate interpretation.  For the
uniform distribution on the chosen finite $U$, Ellerman's logical entropy and
logical mutual information are, respectively,
\begin{equation}
  h(\pi)=\frac{|\dit(\pi)|}{|U|^2},
  \qquad
  m(\pi,\sigma)
  =\frac{|\dit(\pi)\cap\dit(\sigma)|}{|U|^2}.
  \label{eq:logical-entropy}
\end{equation}
In general $m(\pi,\sigma)\ne h(\pi\wedge\sigma)$, because equivalence closure
in Eq.~\eqref{eq:meet-equivalence-closure} gives
\begin{equation}
  \dit(\pi\wedge\sigma)
  \subseteq\dit(\pi)\cap\dit(\sigma),
  \label{eq:mutual-vs-meet}
\end{equation}
possibly strictly.  For the Chinese-lantern partitions,
\begin{equation}
  \dit(\pi)\cap\dit(\sigma)
  =\{(1,4),(4,1),(2,3),(3,2)\},
\end{equation}
so $m(\pi,\sigma)=1/4$, whereas
$h(\pi\wedge\sigma)=h(0_U)=0$ and the shared event algebra is only
$\{\varnothing,U\}$.  Nonzero overlap of distinctions can therefore coexist
with no nontrivial common event proposition
\cite{ellerman-2014-introduction}.  Whether partition implication can organize
Hardy-type forcing chains, or whether these entropy quantities admit a useful
operational reading for a pasted model, remains a representation-dependent
research question rather than a consequence of Ore's correspondence.

\section{Conclusion}

The distinction is between the logic \emph{of} partitions and contextual event
structures \emph{generated by} families of partitions, conventionally called
partition logics.  In the former, whole partitions are classifications or
abstract attributes whose elementary logical instances concern same/different
distinctions between pairs; the partitions admit refinement, join, meet,
implication, and other closure-based operations.  In the latter, partitions
label experimental questions or contexts and their block unions, regarded as
subsets of the hidden-state set, support membership-event propositions in a
pasted event structure.

Ore's correspondence is exact at the level of context generators but does not
identify the two theories.  It maps each $\pi_k$ to $\BA(\pi_k)$, while the
canonical map $q$ sends each event answer to its unordered binary partition
question and identifies complementary answers.  Since $L$ is generally not a
Boolean algebra $\BA(\tau)$, neither construction turns pasting into a
partition connective.  Within a fixed-$U$ representation,
$\BA(\pi\wedge\sigma)=\BA(\pi)\cap\BA(\sigma)$ gives the complete overlap
algebra of a selected context pair, while
$\BA(\pi\vee\sigma)=\langle\BA(\pi)\cup\BA(\sigma)\rangle_{\rm BA}$ gives its
Boolean closure inside $\Pow(U)$.  For the four-state Chinese-lantern
realization, meet $0_U$ records a trivial context intersection and join $1_U$
records the Boolean closure that separates all four hidden states.  More
generally, the bit-string parity comparison shows that the closure can depend
on the chosen concrete realization.  The triangle-shaped family exhibits a
different limitation: its three pairwise meets recover the cyclic shared
outcomes, but its concrete subset order is not the abstract block amalgam and
its literal orthogonal DSDs remain compatible.  More generally, a faithful
set-partition representation requires an order-determining family of
two-valued states; mere separation of event labels need not suffice.  For the Firefly event structure
$L_{12}$, the meet $\{1234,5\}$ records the common binary
classification generated by the shared atom and its complement.  Atomic
intertwining is an additional incidence condition on the original contexts,
expressed by a common block or, in Hilbert space, a common ray.

An orthogonal DSD supplies the corresponding vector-based notion of a
partition: its component projections form a partition of the identity,
orthogonality gives exclusivity, and their sum to $I$ gives completeness.  It
does not partition the set of vectors into equivalence classes; its defining
property is the unique direct-sum decomposition of each vector.

Ellerman's DSD construction generalizes this organization to whole
decompositions or abstract observables and allows incompatible DSDs to lack a
join.  For applications to
Gleason or Kochen--Specker contextuality, the general $\DSD(V)$ framework must
be restricted to orthogonal DSDs, passed to the associated Boolean algebras of
component projections, and supplied with global state constraints.  With that
additional structure, the Firefly incidence pattern is recovered as an event
structure obtained by pasting projection contexts, and state scarcity up to
the Kochen--Specker case can be expressed.  In a concrete fixed-$U$
realization, the same selected partitions can be viewed as whole-partition
objects or as context generators; their carriers, operations, and physical
uses nevertheless remain distinct.

\begin{acknowledgments}
I am grateful to David Ellerman for valuable communications.

OpenAI Codex (GPT-5) was used to assist with literature organization, \LaTeX{}
editing, and checks of derivations.  The authors specified the scientific
direction and prompts; model output retained in the manuscript was checked
against the cited sources and explicit calculations.  The authors accept full
responsibility for the final text.

This research was funded in part by the Austrian
Science Fund (FWF), Grant digital object identifier (DOI)
\href{https://doi.org/10.55776/PIN5424624}{10.55776/PIN5424624}.
The author acknowledges FWF and TU Wien Bibliothek for financial support through its
Open Access Funding Programme.

The authors declare no conflict of interest.
\end{acknowledgments}

\bibliography{svozil}

\begin{thebibliography}{17}%
\makeatletter
\providecommand \@ifxundefined [1]{%
 \@ifx{#1\undefined}
}%
\providecommand \@ifnum [1]{%
 \ifnum #1\expandafter \@firstoftwo
 \else \expandafter \@secondoftwo
 \fi
}%
\providecommand \@ifx [1]{%
 \ifx #1\expandafter \@firstoftwo
 \else \expandafter \@secondoftwo
 \fi
}%
\providecommand \natexlab [1]{#1}%
\providecommand \enquote  [1]{``#1''}%
\providecommand \bibnamefont  [1]{#1}%
\providecommand \bibfnamefont [1]{#1}%
\providecommand \citenamefont [1]{#1}%
\providecommand \href@noop [0]{\@secondoftwo}%
\providecommand \href [0]{\begingroup \@sanitize@url \@href}%
\providecommand \@href[1]{\@@startlink{#1}\@@href}%
\providecommand \@@href[1]{\endgroup#1\@@endlink}%
\providecommand \@sanitize@url [0]{\catcode `\\12\catcode `\$12\catcode
  `\&12\catcode `\#12\catcode `\^12\catcode `\_12\catcode `\%12\relax}%
\providecommand \@@startlink[1]{}%
\providecommand \@@endlink[0]{}%
\providecommand \url  [0]{\begingroup\@sanitize@url \@url }%
\providecommand \@url [1]{\endgroup\@href {#1}{\urlprefix }}%
\providecommand \urlprefix  [0]{URL }%
\providecommand \Eprint [0]{\href }%
\providecommand \doibase [0]{https://doi.org/}%
\providecommand \selectlanguage [0]{\@gobble}%
\providecommand \bibinfo  [0]{\@secondoftwo}%
\providecommand \bibfield  [0]{\@secondoftwo}%
\providecommand \translation [1]{[#1]}%
\providecommand \BibitemOpen [0]{}%
\providecommand \bibitemStop [0]{}%
\providecommand \bibitemNoStop [0]{.\EOS\space}%
\providecommand \EOS [0]{\spacefactor3000\relax}%
\providecommand \BibitemShut  [1]{\csname bibitem#1\endcsname}%
\let\auto@bib@innerbib\@empty
\bibitem [{\citenamefont {Wright}(1990)}]{wright}%
  \BibitemOpen
  \bibfield  {author} {\bibinfo {author} {\bibfnamefont {R.}~\bibnamefont
  {Wright}},\ }\bibfield  {title} {\bibinfo {title} {Generalized urn models},\
  }\href {https://doi.org/10.1007/BF01889696} {\bibfield  {journal} {\bibinfo
  {journal} {Foundations of Physics}\ }\textbf {\bibinfo {volume} {20}},\
  \bibinfo {pages} {881} (\bibinfo {year} {1990})}\BibitemShut {NoStop}%
\bibitem [{\citenamefont {Schaller}\ and\ \citenamefont
  {Svozil}(1995)}]{schaller-95}%
  \BibitemOpen
  \bibfield  {author} {\bibinfo {author} {\bibfnamefont {M.}~\bibnamefont
  {Schaller}}\ and\ \bibinfo {author} {\bibfnamefont {K.}~\bibnamefont
  {Svozil}},\ }\bibfield  {title} {\bibinfo {title} {Automaton partition logic
  versus quantum logic},\ }\href {https://doi.org/10.1007/BF00676288}
  {\bibfield  {journal} {\bibinfo  {journal} {International Journal of
  Theoretical Physics}\ }\textbf {\bibinfo {volume} {34}},\ \bibinfo {pages}
  {1741} (\bibinfo {year} {1995})}\BibitemShut {NoStop}%
\bibitem [{\citenamefont {Dvure{\v{c}}enskij}\ \emph
  {et~al.}(1995)\citenamefont {Dvure{\v{c}}enskij}, \citenamefont
  {Pulmannov{\'{a}}},\ and\ \citenamefont {Svozil}}]{dvur-pul-svo}%
  \BibitemOpen
  \bibfield  {author} {\bibinfo {author} {\bibfnamefont {A.}~\bibnamefont
  {Dvure{\v{c}}enskij}}, \bibinfo {author} {\bibfnamefont {S.}~\bibnamefont
  {Pulmannov{\'{a}}}},\ and\ \bibinfo {author} {\bibfnamefont {K.}~\bibnamefont
  {Svozil}},\ }\bibfield  {title} {\bibinfo {title} {Partition logics,
  orthoalgebras and automata},\ }\href {https://doi.org/10.5169/seals-116747}
  {\bibfield  {journal} {\bibinfo  {journal} {Helvetica Physica Acta}\ }\textbf
  {\bibinfo {volume} {68}},\ \bibinfo {pages} {407} (\bibinfo {year} {1995})},\
  \Eprint {https://arxiv.org/abs/arXiv:1806.04271} {arXiv:1806.04271}
  \BibitemShut {NoStop}%
\bibitem [{\citenamefont {Svozil}(2005)}]{svozil-2001-eua}%
  \BibitemOpen
  \bibfield  {author} {\bibinfo {author} {\bibfnamefont {K.}~\bibnamefont
  {Svozil}},\ }\bibfield  {title} {\bibinfo {title} {Logical equivalence
  between generalized urn models and finite automata},\ }\href
  {https://doi.org/10.1007/s10773-005-7052-0} {\bibfield  {journal} {\bibinfo
  {journal} {International Journal of Theoretical Physics}\ }\textbf {\bibinfo
  {volume} {44}},\ \bibinfo {pages} {745} (\bibinfo {year} {2005})},\ \Eprint
  {https://arxiv.org/abs/arXiv:quant-ph/0209136} {arXiv:quant-ph/0209136}
  \BibitemShut {NoStop}%
\bibitem [{\citenamefont {Ellerman}(2010)}]{ellerman-2010-partition-logic}%
  \BibitemOpen
  \bibfield  {author} {\bibinfo {author} {\bibfnamefont {D.}~\bibnamefont
  {Ellerman}},\ }\bibfield  {title} {\bibinfo {title} {The logic of partitions:
  Introduction to the dual of the logic of subsets},\ }\href
  {https://doi.org/10.1017/S1755020310000018} {\bibfield  {journal} {\bibinfo
  {journal} {The Review of Symbolic Logic}\ }\textbf {\bibinfo {volume} {3}},\
  \bibinfo {pages} {287} (\bibinfo {year} {2010})}\BibitemShut {NoStop}%
\bibitem [{\citenamefont {Ellerman}(2014)}]{ellerman-2014-introduction}%
  \BibitemOpen
  \bibfield  {author} {\bibinfo {author} {\bibfnamefont {D.}~\bibnamefont
  {Ellerman}},\ }\bibfield  {title} {\bibinfo {title} {An introduction to
  partition logic},\ }\href {https://doi.org/10.1093/jigpal/jzt036} {\bibfield
  {journal} {\bibinfo  {journal} {Logic Journal of the IGPL}\ }\textbf
  {\bibinfo {volume} {22}},\ \bibinfo {pages} {94} (\bibinfo {year}
  {2014})}\BibitemShut {NoStop}%
\bibitem [{\citenamefont {Ellerman}(2019)}]{ellerman-2019-graph-partitions}%
  \BibitemOpen
  \bibfield  {author} {\bibinfo {author} {\bibfnamefont {D.}~\bibnamefont
  {Ellerman}},\ }\bibfield  {title} {\bibinfo {title} {A graph-theoretic method
  to define any boolean operation on partitions},\ }\href
  {https://doi.org/10.26493/2590-9770.1259.9d5} {\bibfield  {journal} {\bibinfo
   {journal} {The Art of Discrete and Applied Mathematics}\ }\textbf {\bibinfo
  {volume} {2}},\ \bibinfo {pages} {P2.02} (\bibinfo {year}
  {2019})}\BibitemShut {NoStop}%
\bibitem [{\citenamefont {Ellerman}(2018)}]{ellerman-2018-dsd-quantum-logic}%
  \BibitemOpen
  \bibfield  {author} {\bibinfo {author} {\bibfnamefont {D.}~\bibnamefont
  {Ellerman}},\ }\bibfield  {title} {\bibinfo {title} {The quantum logic of
  direct-sum decompositions: The dual to the quantum logic of subspaces},\
  }\href {https://doi.org/10.1093/jigpal/jzx026} {\bibfield  {journal}
  {\bibinfo  {journal} {Logic Journal of the IGPL}\ }\textbf {\bibinfo {volume}
  {26}},\ \bibinfo {pages} {1} (\bibinfo {year} {2018})}\BibitemShut {NoStop}%
\bibitem [{\citenamefont {Ore}(1942)}]{ore-1942-equivalence}%
  \BibitemOpen
  \bibfield  {author} {\bibinfo {author} {\bibfnamefont {O.}~\bibnamefont
  {Ore}},\ }\bibfield  {title} {\bibinfo {title} {Theory of equivalence
  relations},\ }\href {https://doi.org/10.1215/S0012-7094-42-00942-6}
  {\bibfield  {journal} {\bibinfo  {journal} {Duke Mathematical Journal}\
  }\textbf {\bibinfo {volume} {9}},\ \bibinfo {pages} {573} (\bibinfo {year}
  {1942})}\BibitemShut {NoStop}%
\bibitem [{\citenamefont {Finberg}\ \emph {et~al.}(1996)\citenamefont
  {Finberg}, \citenamefont {Mainetti},\ and\ \citenamefont
  {Rota}}]{finberg-mainetti-rota-1996}%
  \BibitemOpen
  \bibfield  {author} {\bibinfo {author} {\bibfnamefont {D.}~\bibnamefont
  {Finberg}}, \bibinfo {author} {\bibfnamefont {M.}~\bibnamefont {Mainetti}},\
  and\ \bibinfo {author} {\bibfnamefont {G.-C.}\ \bibnamefont {Rota}},\
  }\bibfield  {title} {\bibinfo {title} {The logic of commuting equivalence
  relations},\ }in\ \href {https://doi.org/10.1201/9780203748671-4} {\emph
  {\bibinfo {booktitle} {Logic and Algebra}}},\ \bibinfo {series} {Lecture
  Notes in Pure and Applied Mathematics}, Vol.\ \bibinfo {volume} {180},\
  \bibinfo {editor} {edited by\ \bibinfo {editor} {\bibfnamefont
  {A.}~\bibnamefont {Ursini}}\ and\ \bibinfo {editor} {\bibfnamefont
  {P.}~\bibnamefont {Aglian{\`o}}}}\ (\bibinfo  {publisher} {Marcel Dekker},\
  \bibinfo {address} {New York},\ \bibinfo {year} {1996})\ pp.\ \bibinfo
  {pages} {69--96}\BibitemShut {NoStop}%
\bibitem [{\citenamefont {Svozil}(2020)}]{svozil-2018-b}%
  \BibitemOpen
  \bibfield  {author} {\bibinfo {author} {\bibfnamefont {K.}~\bibnamefont
  {Svozil}},\ }\bibfield  {title} {\bibinfo {title} {Faithful orthogonal
  representations of graphs from partition logics},\ }\href
  {https://doi.org/10.1007/s00500-019-04425-1} {\bibfield  {journal} {\bibinfo
  {journal} {Soft Computing}\ }\textbf {\bibinfo {volume} {24}},\ \bibinfo
  {pages} {10239} (\bibinfo {year} {2020})},\ \Eprint
  {https://arxiv.org/abs/1810.10423} {arXiv:1810.10423} \BibitemShut {NoStop}%
\bibitem [{\citenamefont {Svozil}(2009)}]{svozil-2008-ql}%
  \BibitemOpen
  \bibfield  {author} {\bibinfo {author} {\bibfnamefont {K.}~\bibnamefont
  {Svozil}},\ }\bibfield  {title} {\bibinfo {title} {Contexts in quantum,
  classical and partition logic},\ }in\ \href
  {https://doi.org/10.1016/B978-0-444-52869-8.50015-3} {\emph {\bibinfo
  {booktitle} {Handbook of Quantum Logic and Quantum Structures}}},\ \bibinfo
  {editor} {edited by\ \bibinfo {editor} {\bibfnamefont {K.}~\bibnamefont
  {Engesser}}, \bibinfo {editor} {\bibfnamefont {D.~M.}\ \bibnamefont
  {Gabbay}},\ and\ \bibinfo {editor} {\bibfnamefont {D.}~\bibnamefont
  {Lehmann}}}\ (\bibinfo  {publisher} {Elsevier},\ \bibinfo {address}
  {Amsterdam},\ \bibinfo {year} {2009})\ pp.\ \bibinfo {pages} {551--586},\
  \Eprint {https://arxiv.org/abs/arXiv:quant-ph/0609209}
  {arXiv:quant-ph/0609209} \BibitemShut {NoStop}%
\bibitem [{\citenamefont {Svozil}(2021)}]{svozil-2020-hardy}%
  \BibitemOpen
  \bibfield  {author} {\bibinfo {author} {\bibfnamefont {K.}~\bibnamefont
  {Svozil}},\ }\bibfield  {title} {\bibinfo {title} {Extensions of {H}ardy-type
  true-implies-false gadgets to classically obtain indistinguishability},\
  }\href {https://doi.org/10.1103/PhysRevA.103.022204} {\bibfield  {journal}
  {\bibinfo  {journal} {Physical Review A}\ }\textbf {\bibinfo {volume}
  {103}},\ \bibinfo {pages} {022204} (\bibinfo {year} {2021})},\ \Eprint
  {https://arxiv.org/abs/arXiv:2006.11396} {arXiv:2006.11396} \BibitemShut
  {NoStop}%
\bibitem [{\citenamefont {Harding}(1996)}]{harding-96}%
  \BibitemOpen
  \bibfield  {author} {\bibinfo {author} {\bibfnamefont {J.}~\bibnamefont
  {Harding}},\ }\bibfield  {title} {\bibinfo {title} {Decompositions in quantum
  logic},\ }\href {https://doi.org/10.1090/S0002-9947-96-01548-6} {\bibfield
  {journal} {\bibinfo  {journal} {Transactions of the American Mathematical
  Society}\ }\textbf {\bibinfo {volume} {348}},\ \bibinfo {pages} {1839}
  (\bibinfo {year} {1996})}\BibitemShut {NoStop}%
\bibitem [{\citenamefont {Gleason}(1957)}]{Gleason}%
  \BibitemOpen
  \bibfield  {author} {\bibinfo {author} {\bibfnamefont {A.~M.}\ \bibnamefont
  {Gleason}},\ }\bibfield  {title} {\bibinfo {title} {Measures on the closed
  subspaces of a {H}ilbert space},\ }\href
  {https://doi.org/10.1512/iumj.1957.6.56050} {\bibfield  {journal} {\bibinfo
  {journal} {Journal of Mathematics and Mechanics}\ }\textbf {\bibinfo {volume}
  {6}},\ \bibinfo {pages} {885} (\bibinfo {year} {1957})}\BibitemShut {NoStop}%
\bibitem [{\citenamefont {Kochen}\ and\ \citenamefont
  {Specker}(1967)}]{kochen1}%
  \BibitemOpen
  \bibfield  {author} {\bibinfo {author} {\bibfnamefont {S.}~\bibnamefont
  {Kochen}}\ and\ \bibinfo {author} {\bibfnamefont {E.~P.}\ \bibnamefont
  {Specker}},\ }\bibfield  {title} {\bibinfo {title} {The problem of hidden
  variables in quantum mechanics},\ }\href
  {https://doi.org/10.1512/iumj.1968.17.17004} {\bibfield  {journal} {\bibinfo
  {journal} {Journal of Mathematics and Mechanics}\ }\textbf {\bibinfo {volume}
  {17}},\ \bibinfo {pages} {59} (\bibinfo {year} {1967})}\BibitemShut {NoStop}%
\bibitem [{\citenamefont {Svozil}\ and\ \citenamefont
  {Tkadlec}(1996)}]{svozil-tkadlec}%
  \BibitemOpen
  \bibfield  {author} {\bibinfo {author} {\bibfnamefont {K.}~\bibnamefont
  {Svozil}}\ and\ \bibinfo {author} {\bibfnamefont {J.}~\bibnamefont
  {Tkadlec}},\ }\bibfield  {title} {\bibinfo {title} {Greechie diagrams,
  nonexistence of measures in quantum logics and {K}ochen--{S}pecker type
  constructions},\ }\href {https://doi.org/10.1063/1.531710} {\bibfield
  {journal} {\bibinfo  {journal} {Journal of Mathematical Physics}\ }\textbf
  {\bibinfo {volume} {37}},\ \bibinfo {pages} {5380} (\bibinfo {year}
  {1996})}\BibitemShut {NoStop}%
\end{thebibliography}%

\end{document}